\documentclass[12pt]{article}
\usepackage{endnotes}
\usepackage{osajnl2}

\begin{document}

\title{Resolving the wave-vector and the refractive index from the coefficient of reflectance}

\author{V.~U.~Nazarov$^{*}$ and Y.-C.~Chang}

\address{
Research Center for Applied Sciences, Academia Sinica,  128 Section 2 Academia Road,
Nankang, Taipei 115, Taiwan  \\
$^*$Corresponding author: nazarov@gate.sinica.edu.tw
}

\begin{abstract}
We  resolve  the existing controversy concerning the selection of the sign
of the normal-to-the-interface component of the wave-vector $k_z$ of an electromagnetic wave in an active (gain) medium.
Our method exploits the fact that no ambiguity exists in the case of a {\em film} of the active medium
since its coefficient of reflectance is invariant under the
inversion of the sign of $k_z$. Then we show that the limit of the infinite film thickness determines a unique
and physically consistent choice of the wave-vector and the refractive index.
Practically important implications of the theory are identified and discussed.
\end{abstract}


%] %% activate for two-column option

\maketitle

\begin{figure}[htb]
\centerline{\includegraphics[width=8.5cm]{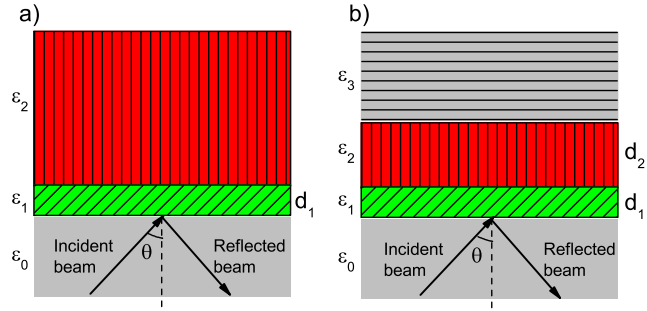}}
\caption{\label{Geometry}
(Color online)
Systems under study: (a)
Semi-infinite dielectric ($\epsilon_0$), metal film ($\epsilon_1$, $d_1$) and a semi-infinite active medium ($\epsilon_2$).
(b) As in (a) but the active medium constitutes a film of the thickness $d_2$, and there is a semi-infinite dielectric on the top with the dielectric function $\epsilon_3$.
}
\end{figure}

\noindent Recently, there has been much of debate
regarding the correct selection of the sign of the normal-to-the-interface component of the
wave-vector $k_z$ of an electromagnetic wave propagating in an active medium
or, equivalently for the normal incidence, of the refractive index of the active medium
\cite{Ramakrishna-05,Chen-05,Mackay-06,Chen-06,Ramakrishna-07,Chen-07}.
Specifically, if one considers a system of a glass prism, a metal film, and an active dielectric
(the Kretschmann-Raether configuration) as schematized in Fig.~\ref{Geometry} (a), then the coefficient of reflectance
is given by \cite{Raether-88} (for the sake of simplicity, we consider nonmagnetic media with $\mu=1$)
\begin{eqnarray}
R= \left| \frac{r_{01}+r_{12} e^{2 i k_{1z} d_1}}{1+r_{01} r_{12} e^{2 i k_{1z} d_1}} \right|^2,
\label{R}
\end{eqnarray}
where the coefficients
\begin{eqnarray}
r_{ij}=(k_{iz} \epsilon_j - k_{jz} \epsilon_i)/(k_{iz} \epsilon_j + k_{jz} \epsilon_i)
\label{rik}
\end{eqnarray}
describe the parallel field components,
\begin{eqnarray}
k_{iz}=\pm \frac{2 \pi}{\lambda} \sqrt{\epsilon_i - \epsilon_0 \sin^2 \theta},
\label{kz}
\end{eqnarray}
$\epsilon_0$, $\epsilon_1$, and $\epsilon_2$ are complex dielectric constants of the prism, the metal,
and the active dielectric, respectively, $\lambda$ is the light wave-length in vacuum, and $\theta$ is the angle of incidence.
For an active medium ($\epsilon_2''<0$) the choice of the sign of the square root in Eq.~(\ref{kz}) is far from evident
\cite{Ramakrishna-05,Chen-05,Mackay-06,Chen-06,Skaar-06,Ramakrishna-07,Chen-07}
while this choice strongly affects the coefficient of reflectance of Eq.~(\ref{R}).

In this Letter, we show that the problem of the sign of $k_{2z}$ can be unambiguously and naturally resolved by first considering
a film of the active dielectric then letting the thickness of the film tend  to infinity. Let us consider a system
in Fig.~\ref{Geometry} (b) which differs from that in Fig.~\ref{Geometry} (a) by the finite thickness $d_2$ of the active dielectric
and the presence of a semi-infinite passive dielectric with the dielectric constant $\epsilon_3$.
For this system the coefficient of reflectance can be written as \cite{footnote}
%\footnote{Eq.~(\ref{RF}) for the reflectance of a four-component system
%can be derived in just the same way as Eq.~(\ref{R}) for a three-component
%system by applying Maxwell's equations and the standard boundary conditions at interfaces.}

\begin{eqnarray}
&&R = \label{RF}    \\
&&\left|\frac{e^{2 i k_{1z}
        d_1} \! \left( r_{12} \! + \!
       e^{ 2 i    k_{2z}
           d_2} r_{23} \right)  \! + \!
    r_{01}
     \left( 1 \! + \! e^
         {2 i   k_{2z} d_2}
          r_{12} r_{23}
       \right) }{1 + e^
      { 2 i    k_{2z} d_2}
     r_{12} r_{23} \! + \!
    e^{ 2 i k_{1z} {d_1}}
     r_{01}
     \left( r_{12} +
       e^{ 2 i    k_{2z}
           {d_2}} r_{23} \right) } \right|^2.
\nonumber
\end{eqnarray}
This is evident [and can also be directly verified with use of Eq.~(\ref{rik})]
that Eq.~(\ref{RF}) is invariant under the transformation $k_{2z}\rightarrow - k_{2z}$.
Therefore, there is no ambiguity in the selection of the sign of $k_{2z}$ in the case of a film of the active dielectric.
Now, putting $d_2$ to infinity in Eq.~(\ref{RF}), we immediately retrieve Eq.~(\ref{R}) if and only if $k_{2z}$ has
a positive imaginary part {\em regardless of whether  medium 2 is active or passive}.
Accordingly, the branch cut in the complex plane of $k_{2z}^2$ must
be taken along the {\em positive part of the real} axis ($0\le \phi < 2\pi$, where $\phi$ is the argument of $k_{2z}^2$),
in agreement with  \cite{Chen-05} and in disagreement with  \cite{Ramakrishna-05} and \cite{Mackay-06,Wei-07}, where the branch cuts were taken along the negative imaginary ($-\pi/2 \le \phi < 3/2 \, \pi)$
and negative real ($-\pi \le \phi < \pi$) axes, respectively.
Therefore, for an active medium
the refractive index $n=\sqrt{\epsilon}$ has negative real part.
This concludes our proof as which  of the two signs of $k_{2z}$ should be chosen  in Eq.~(\ref{kz}) to be substituted into Eq.~(\ref{R}).

\begin{figure}[htb]
\centerline{\includegraphics[width=8.5cm]{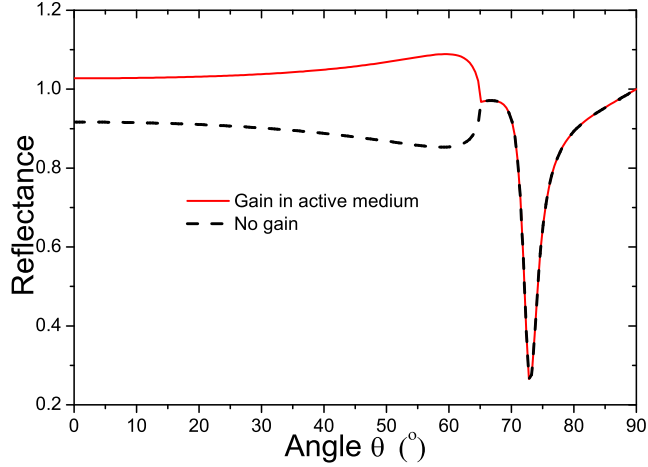}}
\caption{\label{Refl1}
(Color online)
Coefficient of reflectance  of the system of a glass prism, silver film,
and the dye of cresyl violet with ($\epsilon_2= 1.85-9 \times 10^{-6} i$, red solid line)
and without ($\epsilon_2= 1.85 + 0 \times  i$, black dashed line) gain.
Other parameters are those from  \cite{Seidel-05} (see text).}
\end{figure}

\begin{figure}[htb]
\centerline{\includegraphics[width=8.5cm]{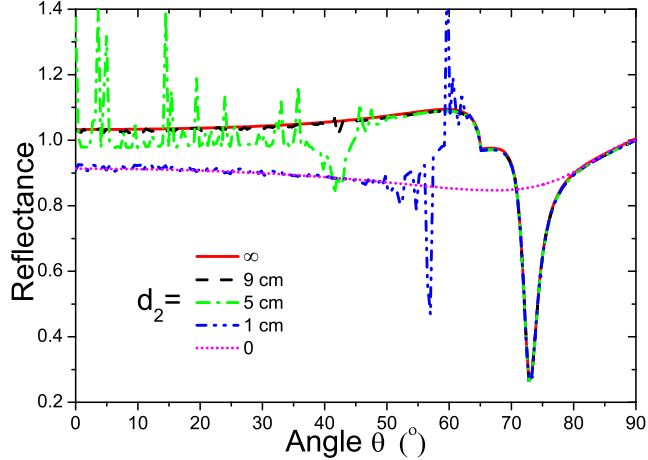}}
\caption{\label{Refl2}
(Color online)
Convergence of the coefficient of reflectance of the system with a finite  film of the active dielectric
to that with the semi-infinite one.
Parameters are those from  \cite{Seidel-05} (see text).}
\end{figure}

\begin{figure}[htb]
\centerline{\includegraphics[width=8.5cm]{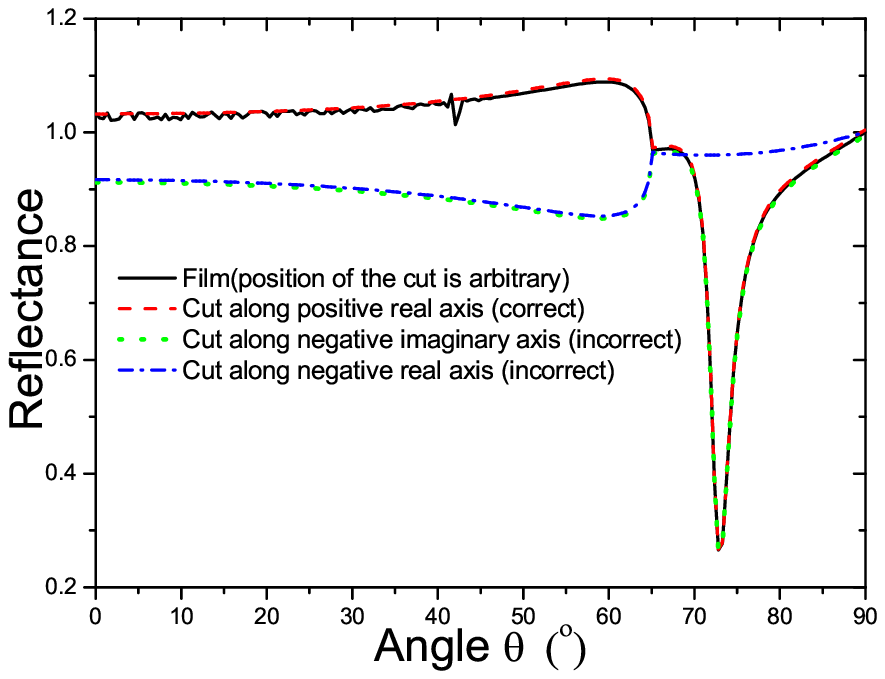}}
\caption{\label{Compare}
(Color online)
Implications of the selection of the three
different branch cuts in $k_z^2$ complex plane.
Red dashed line corresponds to the branch cut along positive real axis
($0\le \phi < 2\pi$).
Green dotted line corresponds to the branch cut along negative imaginary axis
($-\pi/2 \le \phi < (3/2) \pi$) \cite{Ramakrishna-05}.
Blue dashed-dotted line corresponds to the branch cut along negative real axis
($-\pi \le \phi < \pi$) \cite{Mackay-06,Wei-07}.
Black solid line corresponds to a film of the active dielectric ($d_2=9$ cm).
Other parameters are those from  \cite{Seidel-05} (see text).}
\end{figure}

In order to provide an illustrative example which helps to elucidate a number of practically instructive points,
in Figs. \ref{Refl1} -  \ref{Compare} we present numerical results for the coefficient of reflectance of the system of a glass prism, a silver film,
and the dye of cresyl violet.
Parameters as taken from  \cite{Seidel-05} are: $\epsilon_0=2.25$, $\epsilon_1=-18 + 0.7\, i$, $\epsilon_2=1.85-9 \times 10^{-6} i$,
$d_1=39$ nm, and $\lambda=633$ nm.

In Fig. \ref{Refl1} the coefficient of reflectance of Eq. (\ref{R}) of the system with the semi-infinite active dielectric
(red solid line)
is plotted  and compared with that of a passive dielectric (black dashed line).
The sharp dip in the spectrum with the minimum at $73.1^\circ$
is due to the surface plasmon polariton associated with silver film. At lower angles of incidence,
the reflectance is mainly greater than
one for the system with the active dielectric, obviously due to the gain in the latter.
It must be noted that the value of $\epsilon_2''=-9 \times 10^{-6}$ is small enough
that the corresponding spectrum is indistinguishable from that of the system with $\epsilon_2''= 0_-$,
where $0_-$ stands for the zero limit from the negative side. An important point is
that there is no contradiction in reflectance of the systems with $\epsilon_2''= 0_+$ and $\epsilon_2''= 0_-$
to differ finitely (and considerably) as is seen in Fig. \ref{Refl1}, since $\epsilon_2''= 0_-$ combined with {\em infinite}
thickness of the active medium results in the finite overall gain (mathematically, the limit of infinite thickness should be taken first
at finite negative $\epsilon_2''$, then the latter should be put to zero).

In order to understand what films of the active dielectric are thick enough to be considered as semi-infinite,
in Fig. \ref{Refl2} we plot the coefficient of reflectance of Eq. (\ref{RF}) for a number of film thicknesses.
The striking result of the saturation of the spectrum of a film to that of the semi-infinite medium occurring
at the film thickness as huge as about $9$ cm is simply due to the tiny $\epsilon_2''=-9 \times 10^{-6}$,
and this should serve as a warning when weakly active dielectric films are modeled with semi-infinite media.

In Fig. \ref{Compare} we plot the  reflectance of Eq.~(\ref{R})
obtained with three different branch cuts in the complex plane of $k_{2z}^2$.
These are compared with
the coefficient of reflectance of the system with a film of active dielectric ($d_2=9$ cm),
for which the selection of the branch cut does not make a difference,
and which, therefore, at large film thickness can serve as a criterion of the validity of the branch
cut chosen in the semi-infinite case.
In agreement with our formal proof, the spectrum calculated with the branch cut along
the positive real axis (${\rm Im}\, k_z>0$) (red dashed line)
practically coincides at all angles with that of the film (black solid line).
At the same time, the cut along the negative imaginary axis \cite{Ramakrishna-05} (green dashed line) results
in a loss instead of the gain below the critical angle, while the cut along the negative real axis \cite{Mackay-06,Wei-07}
(blue dashed line) leads to the incorrect reflectance in the range of the surface plasmon polariton.

We point out a two-fold significance of deriving  the wave-vector
in a half-space from that for a finite film:
Firstly, this method provides a mathematically rigorous and simple procedure to resolve the sign of the wave-vector.
Secondly, in a more general perspective,
in  experiment there evidently exists no semi-infinite medium
but only (maybe thick) films. A half-space is a convenient abstraction with
properties being the limit of the corresponding properties of a film thick enough
so that a further increase of the film's thickness does not change results of an experiment.
Therefore, our results for semi-infinite active
media are automatically applicable to the interpretation of experimental data with
sufficiently thick films. This is not the case with other choices
of the branch cut in the complex plane of $k_{2z}$ \cite{Ramakrishna-05,Mackay-06,Wei-07} as has been already
discussed in conjunction with Fig.~\ref{Compare}.

Ref.~\cite{Geddes-07} performs a numerical analysis of a wave-packet propagation in a system considered in \cite{Chen-05}.
Correcting errors of  \cite{Mackay-06} by the same authors,
\cite{Geddes-07} reconfirms the conclusion of \cite{Chen-05} that the
refractive index of the active medium considered in the latter reference has a positive imaginary part.
This conclusion is in agreement with results of the present work.
In particular,  \cite{Geddes-07} does not find the pulse refracted into the active medium to grow  unboundedly which perfectly confirms the positive imaginary part of the refractive index.

In conclusion, by considering a semi-infinite active (gain) dielectric
as a limiting case of its finite-thickness film, we have resolved
the existing controversy in determining the normal-to-the-interface component
of the wave-vector $k_z$ and, equivalently, of the sign of the refractive index of the active dielectric.
Specifically, ${\rm Im}\,k_z$ must be nonnegative for both active and passive media, and, accordingly,
the refractive index of an active medium  has a negative real part.
We have shown that this is physical that
the coefficient of reflectance can be discontinuous with respect to
a continuous change of the dielectric constant of a semi-infinite medium from that of a passive to an active one.
We have applied the theory to the system
of a glass prism, silver film, and the weakly active dye of cresyl violet
in Kretschmann-Raether configuration,
having numerically illustrated the above general concepts.


\begin{thebibliography}{10}
\bibitem{footnote}
Eq.~(\ref{RF}) for the reflectance of a four-component system can be derived in
  just the same way as Eq.~(\ref{R}) for a three-component system by applying
  Maxwell's equations and the standard boundary conditions at interfaces.
\end{thebibliography}
\end{document}